\documentclass[12pt,preprint]{aastex}
\usepackage[dvips]{epsfig}
\usepackage[dvips]{rotating}
\bibpunct{(}{)}{;}{a}{}{,}

\newcommand{\he}{HE~1327$-$2326 }
\slugcomment{Revised version, December 2005}

\shorttitle{The Oxygen Abundance of HE~1327$-$2326}
\shortauthors{Frebel et al.}

\begin{document}
\title{The Oxygen Abundance of HE~1327$-$2326\footnote{Based on observations
  collected at the European Southern Observatory, Paranal, Chile (Proposal ID
  075.D-0048).}}

\author{Anna Frebel\altaffilmark{2}, Norbert Christlieb\altaffilmark{3}, John
  E. Norris\altaffilmark{2}, Wako Aoki\altaffilmark{4}, and Martin
  Asplund\altaffilmark{2}}

\altaffiltext{2}{Research School of Astronomy \& Astrophysics, The Australian
National University, Cotter Road, Weston, ACT 2611, Australia;
anna@mso.anu.edu.au, jen@mso.anu.edu.au, martin@mso.anu.edu.au}
\altaffiltext{3}{Hamburger Sternwarte, Gojenbergsweg 112, 21029 Hamburg,
Germany; nchristlieb@hs.uni-hamburg.de} \altaffiltext{4}{National Astronomical
  Observatory of Japan, 2-1-21 Osawa, Mitaka, Tokyo, 181-8588 Japan;
  aoki.wako@nao.ac.jp} 

\begin{abstract}
From a newly obtained VLT/UVES spectrum we have determined the oxygen
abundance of HE~1327$-$2326, the most iron-poor star known to date. UV-OH
lines yield a 1D LTE abundance of $\mbox{[O/Fe]}_{\rm{OH}}=3.7$ (subgiant
case) and $\mbox{[O/Fe]}_{\rm{OH}}=3.4$ (dwarf case). Using a correction of
$-$1.0\,dex to account for 3D effects on OH line formation, the abundances are
lowered to $\mbox{[O/Fe]}=2.8$ and $\mbox{[O/Fe]}=2.5$, respectively, which we
adopt. Without 3D corrections, the UV-OH based abundance would be in
disagreement with the upper limits derived from the O\,I triplet lines:
$\mbox{[O/Fe]}_{\rm{trip}}<2.8$ (subgiant) and $\mbox{[O/Fe]}_{\rm{trip}}<3.0$
(dwarf). We also correct the previously determined carbon and nitrogen
abundances for 3D effects. Knowledge of the O abundance of HE~1327$-$2326 has
implications for the interpretation of its abundance pattern. A large O
abundance is in accordance with HE~1327$-$2326 being an early Population II
star which formed from material chemically enriched by a first generation
supernova. Our derived abundances, however, do not exclude other possibilities
such as a Population III scenario.
\end{abstract}


\section{INTRODUCTION}

\citet{HE1327_Nature} recently reported the discovery of the dwarf or subgiant
HE~1327$-$2326, the most iron-poor star known to date (with
$\mbox{[Fe/H]}_{\rm{NLTE}}=-5.4$). Abundances were derived for nine elements
and upper limits for a further eight \citep{Aokihe1327}, including oxygen
($\mbox{[O/Fe]}<4.0$). No detection of molecular OH lines in the UV was
possible from their Subaru/HDS spectrum. A new attempt is presented here to
measure the oxygen abundance of HE~1327$-$2326 from different O indicators:
UV-OH lines, the [O\,I] at 6300\,{\AA} and the O\,I triplet at 7774\,{\AA},
using a higher quality VLT/UVES spectrum. A measurement of the O abundance of
HE~1327$-$2326 is desired for the investigation into the origin of the star.

As the third most common element in the Universe, O generally is an ideal
tracer of its chemical history. Hence, it has been studied in extensive detail
in metal-poor stars to unravel the earliest evolutionary phases of the Galaxy
which is crucial for an understanding of the formation mechanism of the first
generations of stars. It is not clear how the first low-mass stars could form
in the early Universe. A possibility might involve the C and O yields from
Population III supernova which act as sufficient cooling sources in
star-forming gas clouds producing the first low-mass stars
(e.g. \citealt{UmedaNomotoNature, brommnature}). However, the picture which
emerged from the observational studies is not free from inconsistencies,
making the scientific interpretation difficult. A discrepancy of the O
abundances derived from different O indicators poses a serious, not yet
resolved, problem (for a recent discussion see \citealt{asplund_araa}).

Our new observations of HE1327-2326 are presented in \S 2 and the O abundance
measurements are described in \S 3. We discuss the implications in \S 4.

\section{OBSERVATIONS AND DATA REDUCTION}
Between March and May 2005, HE~1327$-$2326 was observed with the
Ultraviolet-Visual Echelle Spectrograph \citep{Dekkeretal:2000} at the Very
Large Telescope, Chile. For the service mode observations we made use of the
dichroic mode and three wavelength settings. The total exposure time of 18\,h
was divided into 18 one hour exposures with the BLUE 346\,nm setting covering
3050--3870\,{\AA}, 15 simultaneous one hour exposures with the RED 580\,nm
setting covering 4780--6805\,{\AA}, and 3 simultaneous one hour exposures with
the RED 760\,nm setting covering 5720--9470\,{\AA}. A $1''$ slit width was
used in the blue arm of the spectrograph, yielding a resolving power of
$R\sim46,000$ while a $0.6''$ slit width was used in the red arm, yielding
$R\sim70,000$. All data have been reduced with the \texttt{REDUCE} package
\citep{reduce}. Overlapping echelle orders were subsequently merged, and the
resulting spectra rebinned to an appropriate sampling. A signal-to-noise of
$S/N\sim40$ was estimated at $\sim3110$\,{\AA}. To ensure the detection of
weak features, unmerged individual orders were used for their verification.

\section{THE OXYGEN ABUNDANCE}
\subsection{The Model Atmosphere}
We performed a 1D LTE abundance analysis of the newly acquired VLT/UVES
spectrum. The latest version of the MARCS code\footnote{Numerous models for
different stellar parameters and compositions are readily available at
http://marcs.astro.uu.se} (Gustafsson et al., in preparation) was used to
compute a model tailored to the chemical abundances observed in HE~1327$-$2326
based on the subgiant abundances reported in \citet{HE1327_Nature}.
Furthermore, we adopted their effective temperature of \mbox{T$_{\rm
eff}=6180$}\,K as well as the two solutions for the surface gravity, $\log
g=3.7$ (subgiant) and $\log g=4.5$ (dwarf). For more details and the
derivation of the stellar parameters of HE~1327$-$2326, we refer the reader to
\citet{Aokihe1327}. For the OH analysis we used the \citet{gillis_ohlinelist}
line list. For the O\,I triplet lines we used data taken from the NIST
database. The $\log gf$ value for the resonance [O\,I] line was taken from
\citet{storey}.

To confirm the validity of our abundance determination technique we took a
MARCS model with the stellar parameters matching those of the well-studied
subgiant HD140283 (we adopt \mbox{T$_{\rm eff}=5850$}\,K, $\log g=3.6$,
$\mbox{[Fe/H]}=-2.46$; see \citealt{boesgaard99} for details). We computed
synthetic spectra with different O abundances to reproduce the Boesgaard et
al. spectrum of HD140283 in the UV-OH line region around $3135$\,{\AA}. From
the comparison of the synthetic with the observed spectrum we derive an
abundance of $\mbox{[O/Fe]}=1.1\pm0.2$ which is in good agreement with the
results derived by \citet{boesgaard99} ($\mbox{[O/Fe]}=1.05$) using the solar
abundance adopted by them. From the O triplet lines we derived
$\mbox{[O/Fe]}=0.5$ which also agrees well with the \citet{boesgaard99} value
($\mbox{[O/Fe]}=0.6$).

\subsection{The 1D LTE Analysis}
In our new spectrum we detect 12 Fe\,I lines in the UV spectral range. Seven
of those have already been detected in the Subaru spectrum of our previous
analysis (see \citealt{Aokihe1327} for more details). Our LTE metallicity
derived from these lines is $\mbox{[Fe/H]}_{\rm{LTE}}=-5.7\pm0.2$ for both the
subgiant and dwarf solution. Employing the same NLTE correction as in
\citet{HE1327_Nature} results in an iron abundance for HE~1327$-$2326 in good
agreement with the previously reported metallicity. Unfortunately it is not
possible to detect any Fe\,II lines. Our upper limits are
$\mbox{[Fe\,II/H]}_{\rm{LTE}}<-5.4$ (subgiant) and
$\mbox{[Fe\,II/H]}_{\rm{LTE}}<-5.2$ (dwarf) which are significantly tighter
than the previous values of $\mbox{[Fe\,II/H]}_{\rm{LTE}}<-4.4$ and
$\mbox{[Fe\,II/H]}_{\rm{LTE}}<-4.1$, respectively \citep{Aokihe1327}.

\begin{figure}[ht]
\includegraphics[clip=true]{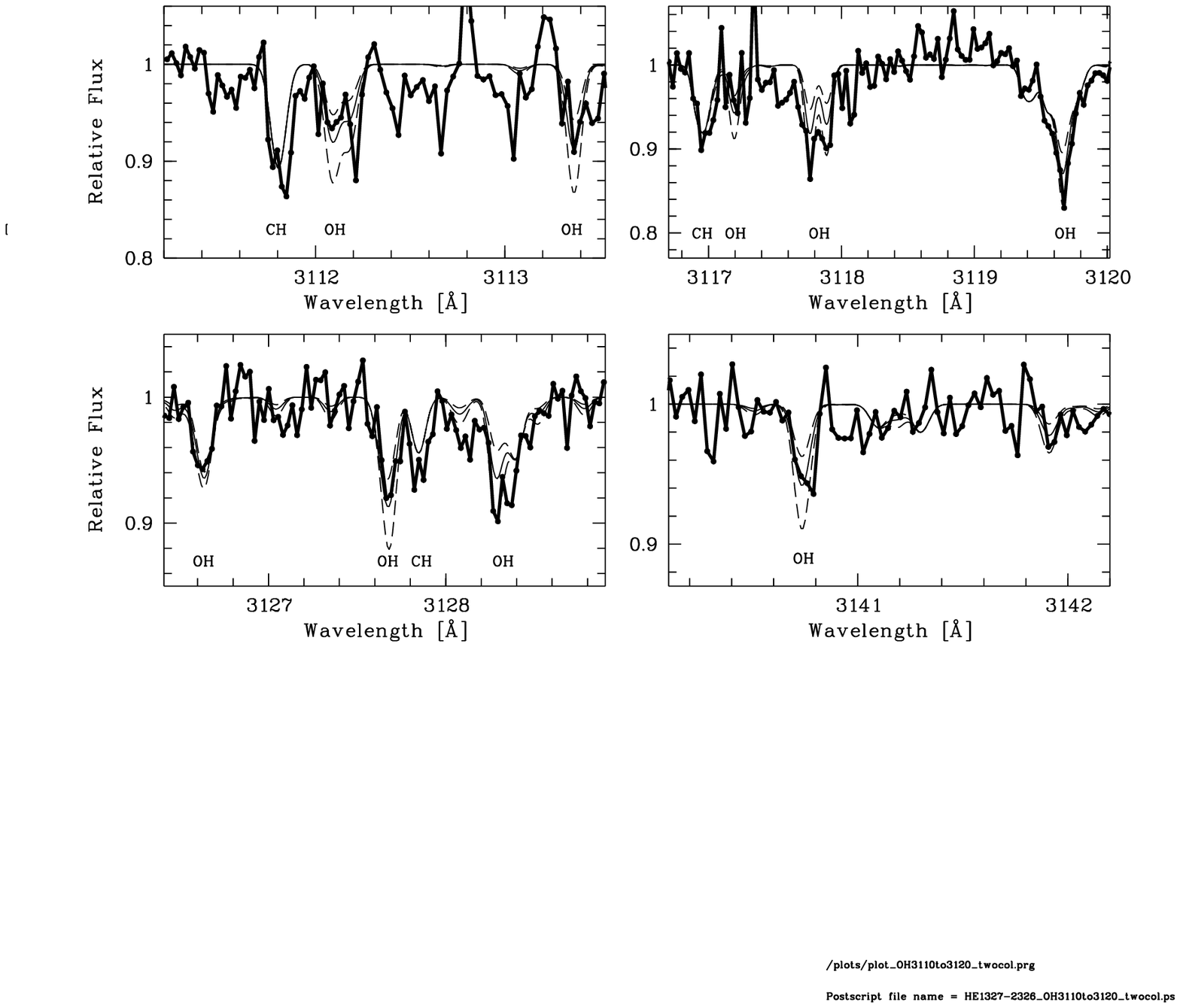}
\figcaption{\label{OH_plot} OH molecular features detected in the UV spectrum
  of HE~1327$-$2326 (thick line). The synthetic 1D LTE spectra for the subgiant
  solution are overplotted (thin and dashed lines) for abundances of
  $\mbox{[O/Fe]}=3.5,3.7,3.9$. The strongest OH and CH features are
  indicated. We used a carbon abundance of $\mbox{[C/Fe]}=4.0$.}
   \end{figure}

Molecular lines from the OH \mbox{$A\;^{2}\Sigma-X\;^{2}\Pi$} system in the
ultraviolet range of our UVES spectrum are clearly detected. Examples are
presented in Figure \ref{OH_plot}. A spectrum synthesis analysis of eight of
the most prominent OH features between 3110 and 3142\,{\AA} was performed. We
note that many CH lines are present over the entire UV spectral range. In this
region, however, the strongest OH lines are visible and some are not strongly
contaminated by CH features. Lines that are as free as possible from such
contamination were used. The total number of OH features, however, is small
and additionally they are in some instances very weak, so that this attempt
was hampered in a few cases. To account for these contaminations we
re-determined the C abundance from UV CH $C-X$ lines around $3180$\,{\AA} (see
Table~\ref{results}). The newly derived values are consistent with our
previous 1D LTE measurements presented in \citet{HE1327_Nature} and
\citet{Aokihe1327}: $\mbox{[C/Fe]}=4.1$ (subgiant) and $\mbox{[C/Fe]}=3.9$
(dwarf) based on CH feature from the $A-X$ and $B-X$ system. A set of
synthetic spectra was computed for a variety of O abundances with a C
abundance set to our new value. Abundances were measured from several OH lines
comparing the observed normalized spectrum with a set of synthetic spectra and
minimizing the $\chi^{2}$ between the synthetic and observed spectrum. We
adopt the average abundance as derived from the individual fits to the OH
features as our final 1D LTE O abundance. The dispersion of the individual
abundance measurements is 0.3\,dex, and the standard error of the mean is
0.1\,dex. However, systematic uncertainties could arise from continuum
placement and, more significantly, from any error in the effective temperature
and the gravity. Taking these uncertainties into account we estimate a total
error of $0.2\,$dex. Thus, $\mbox{[O/Fe]}_{\rm{OH}}=3.7\pm0.2$ (subgiant) and
$\mbox{[O/Fe]}_{\rm{OH}}=3.4\pm0.2$ (dwarf) were adopted as the final averaged
1D LTE abundances. These values are consistent with the upper limit for O of
$\mbox{[O/Fe]}_{\rm{OH}}<4.0$ reported in \citet{HE1327_Nature}. For the solar
O abundance we adopt $\log \epsilon(\rm{O})_{\odot}=8.66$ \citep{solar_abund}.

Considering the very high overabundance derived from the OH features one might
anticipate a detection of the O\,I triplet lines at 7772, 7774 and
7775\,{\AA}. However, despite very high quality data ($S/N\sim260$ at
$\sim7770$\,{\AA}) none of the three lines was detected. Using the formula
$\sigma=w\times \sqrt{n_{pix}}/(S/N)$ (where $w$ is the pixel width, $n_{pix}$
the number of pixels across the line and $S/N$ per pixel; \citealt{bohlin}) we
calculate a $3\sigma$ upper limit ($W_{\lambda}<2$\,m{\AA}) for the strongest
triplet line (7772\,{\AA}). The abundance obtained from this equivalent width
estimate is significantly lower than the abundance derived from OH (employing
the 1D LTE analysis).  Unfortunately, the upper limit (derived for a line
strength limit $W_{\lambda} <1.8$\,{\AA}) for the forbidden [O\,I] line at
6300\,{\AA} is quite large and therefore has little meaning.
The abundances and upper limits can be found in Table~\ref{results}.

\subsection{Application of 3D and NLTE Corrections} 
From theoretical work which investigates beyond the ``classical'' 1D analysis
it is known that the 1D LTE abundance derived from the OH features in
metal-poor stars is significantly higher than the 3D counterpart
\citep{ohnlte_asplund}. This is particularly the case for stars close to the
turnoff. The O triplet lines in turn mostly suffer from 1D NLTE effects
(e.g. \citealt{kiselman93}). For forbidden lines (e.g. [O\,I]), the LTE
assumption is valid and 3D LTE effects are expected to be relatively minor
\citep{nissen02}. The observational discrepancy of O abundances derived from
different indicators are eased with the application of such 3D and/or NLTE
corrections \citep{ohnlte_asplund}. Unfortuntely, appropriate corrections are
not always available.  For the O\,I triplet lines we computed new NLTE
corrections for HE~1327$-$2326 (without consideration of inelastic H
collisions). The 1D NLTE abundance corrections are $-0.3$\,dex for both the
subgiant and dwarf case. We note that 3D NLTE calculations for O\,I in
metal-poor stars with parameters appropriate for \he are not yet available but
preliminary investigations for other halo stars reveal similar 3D NLTE effects
as in 1D \citep{asplund_araa}. See Table~\ref{results} for the corrected
abundances.

\citet{ohnlte_asplund} investigated the difference of 3D LTE model compared to
standard 1D LTE analysis using UV-OH lines. In their Table~2, they provide
corrections for a small set of stars with different stellar parameters, one
set of which is close to that of HE~1327$-$2326. For the two lines
investigated the correction is $-1.0$\,dex for the abundance derived from the
3139.17\,{\AA} OH line and $-0.9$\,dex from the 3167.17\,{\AA} OH line. Since
both lines are too weak to be detectable in HE~1327$-$2326, we simply adopt
the average of the two corrections and apply a $-$1.0\,dex 3D correction to
our 1D LTE O abundance from OH. The abundances from UV-OH lines are thus
lowered significantly and agree well with the upper limits derived from O
triplet lines. NLTE effects on the abundance derived from the OH lines have
not been studied in detail.  We wish to caution here that since HE~1327$-$2326
has a much lower iron abundance than the 3D model from which we inferred the
adopted 3D LTE corrections, it is possible that the real 3D correction might
even be larger than the $-1.0$\,dex applied here.
However, no calculations tailored for the specific abundances of
HE~1327$-$2326 are currently available to further test this assumption.

The formation of molecular CH and NH features is likely to be very similar to
those of OH. \citet{asplund_araa} computed corrections for turnoff stars with
$\mbox{[Fe/H]}=-3.0$ of $-0.6$\,dex for the C abundance and $-0.9$\,dex for
the N abundance. For completeness we thus apply the 3D corrections to the 1D
abundances derived from CH and NH. See Table~\ref{results} for the 3D
corrected abundances. However, any effects would tend to cancel out if the
ratio of any of those elements is to be used.  Similarly large 3D corrections
have recently been computed \citep{colletIAU} for HE~0107$-$5240
\citep{HE0107_Nature}. 

\section{DISCUSSION}
\subsection{Implications of a High Oxygen Abundance}
In order to learn about the earliest stages of star formation in the Universe
it is very important to identify the origin of the elements observed in
HE~1327$-$2326. Oxygen is a key element in this quest because it provides
strong constraints on the different origin scenarios previously invoked for
the star. Of particular importance is whether HE~1327$-$2326 is an early
Population II or a Population III star. Recently, \citet{iwamoto_science} made
an attempt to explain the abundance pattern of HE~1327$-$2326. They invoke a
pre-enrichment scenario in which a faint $25\,M_{\odot}$ Population~III
supernova undergoes a mixing and fallback process producing ejecta containing
little iron and large amounts of CNO. Based on the 1D LTE abundances of
HE~1327$-$2326 and constrained by the 1D LTE upper limit of oxygen
\citep{HE1327_Nature} they compute an O abundance of $\mbox{[O/Fe]}\sim4.0$
which is close to our 1D~LTE abundance derived from OH lines. However, our
adopted 3D abundance is significantly lower. It remains to be seen if their
model could also reproduce our new CNO values since it might be difficult to
simultaneously fit a lower O together with e.g. the high Mg abundance
(potential 3D corrections for Mg are expected to be less severe than for
OH). 

\citet{meynet05} predict a similarly high oxygen abundance
($\mbox{[O/Fe]}=3.5$) based on their combined stellar wind and supernova
ejecta of their rotating $\mbox{[Fe/H]}=-6.6$ stellar models. This is in
qualitative agreement with the observed excesses of O in HE~1327$-$2326 and
other metal-poor stars.  

\begin{figure}[ht]
\includegraphics[clip=true]{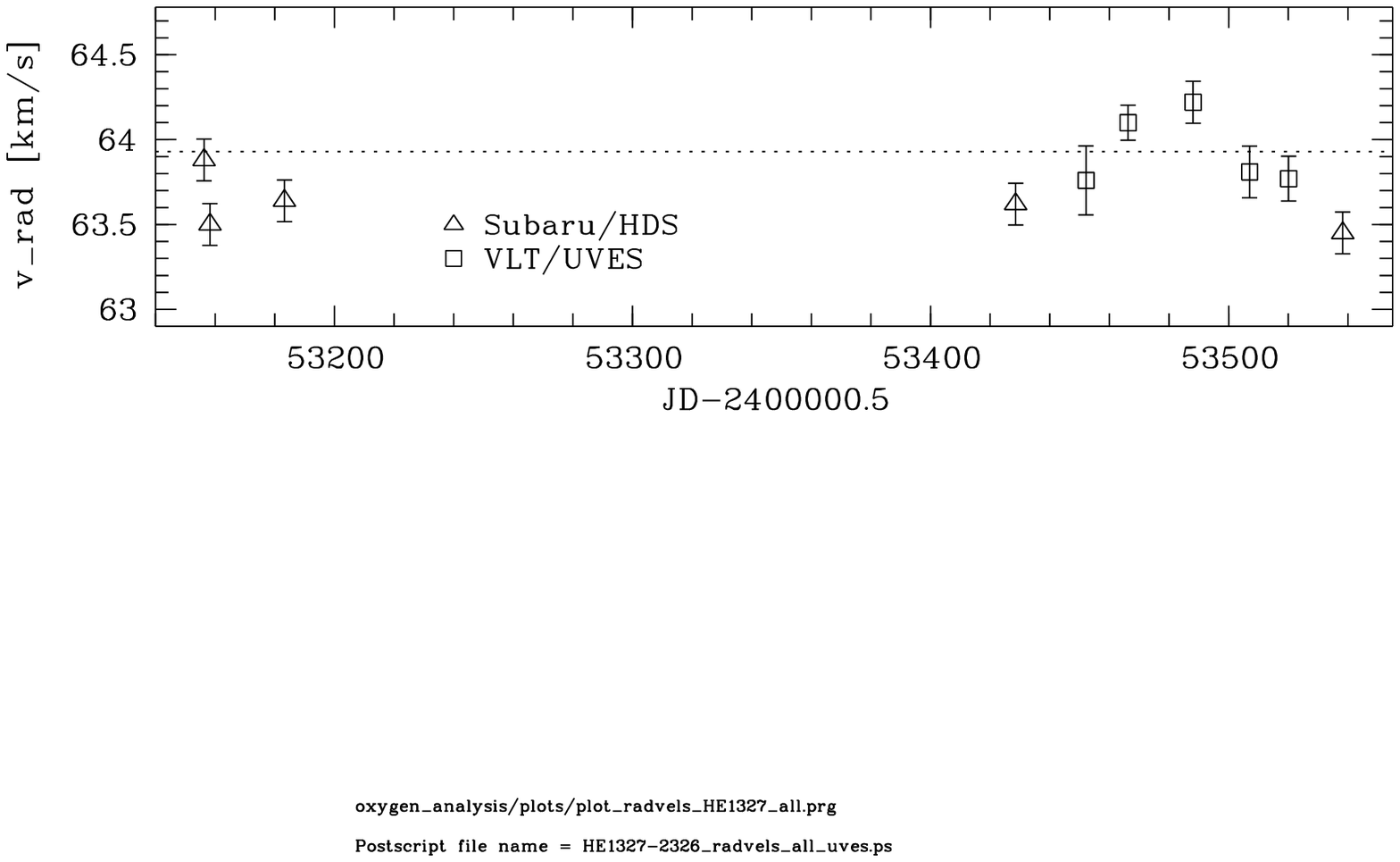} \figcaption{\label{radvel} Barycentric
  radial velocity measurements of HE~1327$-$2326 over a period of 383
  days. The data labeled ``Subaru'' have been taken from Aoki et al. 2005. The
  dotted line presents the average of all measurements of 63.9\,km/s. The
  error bars indicate the standard error of the averaged velocities measured
  in observations taken during a night. Given the estimated total measurement
  uncertainty of 0.7--1.0\,km/s which includes systematic errors due to
  instrument instabilities no significant radial velocity variations have
  been detected so far.
}
\end{figure}

Following \citet{suda}, a Population~III scenario might explain the origin of
HE~1327$-$2326 in terms of a binary system. It would then have accreted its
heavier elements from the interstellar medium and the lighter elements from an
erstwhile AGB companion in a binary system. However, the absence of
significant radial velocity variations (see Figure \ref{radvel}) over a period
of just over one year does not support this idea. Within the overall error
there is no change to report in the radial velocity so far. The slight offset
between the Subaru and UVES data points in Figure \ref{radvel} can be
accounted to uncertainties in the wavelength calibrations.  Further work is
required to ascertain whether the O abundance of \he can be explained in this
manner.

Despite the uncertainties of the corrections to the 1D LTE analysis, it is
clear that HE~1327$-$2326 belongs to the group of stars displaying very large
CNO abundances. It appears that the majority of these objects have very low
metallicities (i.e. $\mbox{[Fe/H]}<-3.0$) and that HE~1327$-$2326 is the most
extreme example of the group. However, HE~1327$-$2326 has a similar overall
CNO abundance pattern compared to the only other known star having
$\mbox{[Fe/H]}<-5.0$, HE~0107$-$5240 \citep{HE0107_Nature, O_he0107}. The
unusually high excesses of O of these objects underline that there is no
defined trend amongst the stellar O abundances at the lowest
metallicities. This suggests that there might not be a simple explanation for
the origin of O in the very earliest phases of the Galaxy.

\subsection{Concluding Remarks}
In summary, we adopt the final O abundance to be $\mbox{[O/Fe]}=2.8\pm0.2$
(subgiant) or $\mbox{[O/Fe]}=2.5\pm0.2$ (dwarf). These values are consistent
with the upper limits derived from the O\,I triplet at $\sim 7775$\,{\AA} and
the [O\,I] line at 6300\,{\AA}. This would not be the case if the 1D LTE
abundances derived from OH had been adopted. We note here that atomic
diffusion might have modified the abundances of HE~1327$-$2326. According to
theoretical calculations of \citet{richard2002}, the O/Fe ratio might
originally have been higher. However, observational confirmation of
their calculations is still pending. In any case HE~1327$-$2326 provides
strong observational evidence that 3D LTE effects for the O abundances derived
from OH lines using 1D LTE model atmospheres have to be taken into account,
especially for hotter metal-poor stars.  Where already available, such
corrections should generally be applied when deriving O abundances for
metal-poor stars. A systematic investigation of newly corrected O abundances
with respect to metallicity is clearly desirable.

The newly derived O abundance provides additional constraints on the
Population II models proposed for HE~1327$-$2326 and other metal-poor
stars. Whether or not the new abundance can be reproduced by those models
remains to be seen. Finally we wish to mention that the
\citet{iwamoto_science} model does not include neutron-capture elements. Thus
it is not clear whether the high Sr abundance in HE~1327$-$2326 could be
accounted for with their model. However, recent computations by
\citet{froehlich} indicate that a Sr excess could be in agreement with the
faint SN scenario of Iwamoto et al. We note too that the absence of Li lacks
explanation. The Population III binary scenario might account for the low Li
abundance \citep{HE1327_Nature, Aokihe1327}, but radial velocity variations
have not yet been detected. Hence, a longer time span is needed to monitor the
star for such variations in order to draw a final conclusion. In the absence
of such data we favor the Population II interpretation of HE~1327$-$2326.

\acknowledgments We thank K. Eriksson for computing a tailored MARCS model for
us and A. Korn for helpful comments. We express our gratitude to the ESO staff
on Paranal for carrying out the observations with
VLT-UT2. A.F. thanks N. Piskunov for help with the data reduction and
acknowledges generous hospitality by the Uppsala Astronomical Observatory
where the reduction was carried out. A.F., J.E.N. and M.A. acknowledge support
from the Australian Research Council under grant DP0342613 and N.C. from
Deutsche Forschungsgemeinschaft under grants Ch~214/3 and Re~353/44. This
research has made use of the NIST atomic database, operated by the National
Institute of Standards and Technology.

{\it Facility:} \facility{VLT:Kueyen(UVES)}.

\newpage


\begin{deluxetable}{lrrrrrrr}
\tablecolumns{8}
\tablewidth{0pc}
\tablecaption{\label{results}
Derived abundances of Fe, C and O as well as upper limits of additional
O indicators in HE~1327$-$2326} 
\tablehead{
\colhead{} & \multicolumn{3}{c}{Subgiant} && \multicolumn{3}{c}{Dwarf}\\
\cline{2-4} \cline{6-8} \\
\colhead{Feature}&
\colhead{$\log\epsilon\mbox{(X)}$}&\colhead{$\mbox{[X/H]}$}& \colhead{$\mbox{[X/Fe]}$\tablenotemark{a}}& \colhead{}&
\colhead{$\log\epsilon\mbox{(X)}$}&\colhead{$\mbox{[X/H]}$}& \colhead{$\mbox{[X/Fe]}$\tablenotemark{a}}}
\startdata
Fe\,I LTE   &  1.7 & $-$5.7 &\nodata&& 1.7 & $-$5.7& \nodata \\
Fe\,I NLTE  &  1.9 & $-$5.5 &\nodata&& 1.9 & $-$5.5& \nodata \\
CH LTE (1D) &  6.9 & $-$1.5 & 4.0   && 6.6 & $-$1.8& 3.7 \\
CH LTE (3D) &  6.3 & $-$2.1 & 3.4   && 6.0 & $-$2.4& 3.1 \\
\cline{1-8}\\
OH LTE (1D) &  6.8 & $-$1.8 & 3.7    &&   6.5 & $-$2.1 &   3.4\\
OH LTE (3D) &  5.9 & $-$2.8 & 2.8    &&   5.6 & $-$3.1 &   2.5\\
O\,I LTE    &$<$6.2&$<-$2.5 &$<$3.1  &&$<$6.5 &$<-$2.2 &$<$3.3\\ 
O\,I NLTE   &$<$5.9&$<-$2.7 &$<$2.8  &&$<$6.2 &$<-$2.5 &$<$3.0\\
$\mbox{[O\,I]}$&$<$8.0&$<-$0.7&$<$4.9&&$<$8.3 &$<-$0.4 &$<$5.1
\enddata
\tablecomments{Correcting the 1D LTE nitrogen abundance of HE~1327$-$2326
(\citealt{Aokihe1327}; $\mbox{[N/Fe]}=4.6$ (subgiant), $\mbox{[N/Fe]}=4.1$
(dwarf)) for 3D effects results in $\mbox{[N/Fe]}_{\rm{3D}}=3.7$ (subgiant) and
$\mbox{[N/Fe]}_{\rm{3D}}=3.2$ (dwarf). See text for discussion. }
\tablenotetext{a}{For abundances $\mbox{[X/Fe]}$ we consistently used the NLTE
corrected Fe abundance.  An error of $\pm0.2$\,dex was estimated for all
abundances.}
\end{deluxetable}

\clearpage

\end{document}